\documentclass[preprint2]{aastex}
\usepackage{graphicx}
\usepackage{natbib}

\def\lsim{\mathrel{\rlap{\lower 3pt \hbox{$\sim$}} \raise 2.0pt \hbox{$<$}}}
\def\gsim{\mathrel{\rlap{\lower 3pt \hbox{$\sim$}} \raise 2.0pt \hbox{$>$}}}

\newcommand{\Pbox}{\ensuremath{P_{\rm box}}}
\newcommand{\msun}{\ensuremath{M_\odot}}

\newcommand{\msf}{\ensuremath{M_{\rm SF}}}

\shorttitle{Resolved stellar populations with ELTs}
\shortauthors{Greggio et al.}

\begin{document}

\title{Resolved stellar population of distant galaxies in the ELT era.}

\author{L. Greggio,\email{laura.greggio@oapd.inaf.it} R. Falomo, S. Zaggia and D. Fantinel} 
\affil{INAF - Osservatorio Astronomico di Padova, Vicolo dell'Osservatorio 5,
             I-35122, Padova, Italy}
\and
\author{M. Uslenghi}
\affil{INAF - Istituto di Astrofisica Spaziale e Fisica Cosmica, Via
Bassini 15, I-20133 Milano, Italy}

%
%\author[Greggio et al.]{
%L. Greggio$^{1}$\thanks{E-mail: laura.greggio@oapd.inaf.it}, 
%R. Falomo$^{1}$, 
%S. Zaggia$^{1}$, 
%D. Fantinel$^{1}$, 
%M. Uslenghi$^{2}$
%\\

%\date{Accepted ... Received ... ; in original form ...}

%\pagerange{\pageref{firstpage}--\pageref{lastpage}} \pubyear{2009}

\begin{abstract}
The expected imaging capabilities of future Extremely Large Telescopes
(ELTs) will  offer the unique  possibility to investigate  the stellar
population of  distant galaxies  from the photometry  of the  stars in
very crowded fields.  Using  simulated images and photometric analysis
we explore  here two representative science cases  aimed at recovering
the characteristics of the stellar populations in the inner regions of
distant galaxies. Specifically: case~A) at the center of the  disk of a giant
spiral in the Centaurus Group,  ($\mu_{\rm B}\sim$ 21, distance of 4.6
Mpc); and, case~B) at  half  of  the  effective  radius of  a  giant
elliptical in the Virgo Cluster  ($\mu_{\rm I} \sim$ 19.5, distance of
18 Mpc). We generate synthetic frames by distributing model
stellar populations and adopting a representative instrumental set up,
i.e. a  42 m Telescope operating close to the diffraction limit. 
The  effect  of  crowding is  discussed  in detail showing how
  stars are measured preferentially brighter than they are as
  the confusion limit is approached.
We  find that (i) accurate photometry ($\sigma
\sim 0.1$,  completeness $\gsim 90$\%) can be obtained  for case~B)
down  to $I\sim 28.5$,  $J\sim 27.5$  allowing us  to recover  the
stellar metallicity  distribution in the inner  regions of ellipticals
in Virgo to within $\sim 0.1$~dex; (ii) the same photometric accuracy
holds  for the  science case~A) down  to $J\sim 28.0$,  $K\sim 27.0$,
enabling to reconstruct of  the star formation history up to
the Hubble time  via simple star counts in  diagnostic boxes. For this
latter  case we  discuss  the possibility  of  deriving more  detailed
information on the  star formation history from the  analysis of their
Horizontal Branch  stars. We show  that the combined features  of high
sensitivity and angular resolution of ELTs  may open a new era for our
knowledge   of  the   stellar   content  of   galaxies  of   different
morphological type up to the distance of the Virgo cluster.
\end{abstract}


% --------------------------------------------------------------------------------
\section{Introduction}

One of the key issues  in modern astronomy concerns the Star Formation
History (SFH) in  the Universe. Direct observations of  galaxies up to
high redshift  can be used  to map the  SFH, but since  the integrated
galaxy  light is  dominated by  the  contribution of  the most  recent
generations  of stars,  the information  that  can be  derived on  the
underlying older stellar population  is severely limited. However, the
SFH in galaxies can  be uncovered by interpreting the Colour-Magnitude
Diagrams  (CMD) of  their stars,  which contain  the fossil  record of
their  SFH (e.g., \citealt{greggio}, \citealt{holtz}, \citealt{tom},
\citealt{cole}, \citealt{mcq}, \citealt{weisz}; see also
\citealt{cigno}  for a  review). 
This  kind of
studies  require  accurate  photometry  down to  faint  magnitudes  in
crowded fields, which, with  current instrumentation, is feasible only
for the nearest galaxies. This  implies a very limited sampling of the
SFH in the Universe, with plenty of dwarfs, a few spirals and no giant
elliptical.

By  the  end of  this  decade this  situation  is  expected to  change
significantly, as a number of extremely large telescopes, as the Giant
Magellan Telescope (GMT;  \citealt{johns}), the Thirty Meter Telescope
(TMT;  \citealt{szeto})  and the  European  Extremely Large  Telescope
(E-ELT;  \citealt{gilmozzi})  could come  into  operation.  The  large
collecting area of  ELTs coupled with adaptive optics  cameras able to
deliver quasi-diffraction limited images will allow us to probe a wide
volume, where we  can access a significative sample  of galaxies. More
importantly,  it  will be  possible  to  study  dense stellar  fields.
Indeed, because  of crowding, stellar photometry  in external galaxies
is  currently  feasible only  in  regions  of  relatively low  surface
brightness.  For  giant  galaxies,  this  prevents  us  from  deriving
detailed  SFH where  most  of the  galaxy  mass is.   One  of the  key
advantages of  taking images with extremely  large aperture telescopes
is the  exceptionally good image  quality when the telescope  can work
close to  its diffraction limit. This significant  improvement has two
fundamental  effects :  1) it  produces  a dramatic  reduction of  the
background light over the  point spread function (PSF) area considered
for  the photometry;  2) it  allows a  significant improvement  in the
spatial  resolution (proportional  to the  telescope  aperture). These
combined  advantages  offer  the   unique  opportunity  to  carry  out
observations  of faint  targets in  crowded and/or  structured objects
(galaxies) at  large distance, that  cannot be exploited by  any other
telescope  of smaller  aperture, neither  ground nor  space  based. In
summary, the excellent resolution capabilities of next generation
large aperture Telescopes will enable us
to  study high  surface brightness  regions, i.e.  the inner  parts of
galaxies, where  star formation  was more conspicuous,  as well  as to
address directly age and metallicity gradients.

The photometric performance  of ELTs operating close to the
diffraction   limit   was   investigated   by  \cite{olsen}   and   by
\cite{atul}.  Both studies focus on the impact of 
crowding conditions on the photometric quality, as
quantified by the 1 $\sigma$ width of the error distribution as a
function of magnitude, for stellar fields with different surface
brightness. In this approach errors are considered to be
  symmetric, with stellar luminosities having the same probability of
  being overestimated or underestimated. However, crowding 
induces  asymmetrical photometric  errors, with an excess of stars
  measured brighter than they are (e.g. \citealt{tosi};
    \citealt{carma96}; \citealt{alviopix}). This results into artificially
  brightened features on the CMD which may induce a systematic
    error in those parameters (e.g. distances and stellar ages) which are derived from
  their luminosity.  In addition, crowding may also affect 
  systematically the distribution of stars across the
  CMD, jeopardizing its interpretation in terms of star formation
  history. The impact of this asymmetry is negligible under
    low-crowding conditions (e.g. when the 1$\sigma$ width is smaller
    than about 0.1 mag), but grows rapidly with crowding.
In this paper we present the results of end-to-end simulations
of two specific science cases aimed at
investigating   the  SFH   in  galaxies,   which  fully   exploit  the
unprecedented capabilities foreseen for ELTs.

% --------------------------------------------------------------------------------
\begin{figure}[t]
%\resizebox{\hsize}{!}{\includegraphics{limits.eps}}
%\plotone{limits.eps}
\plotone{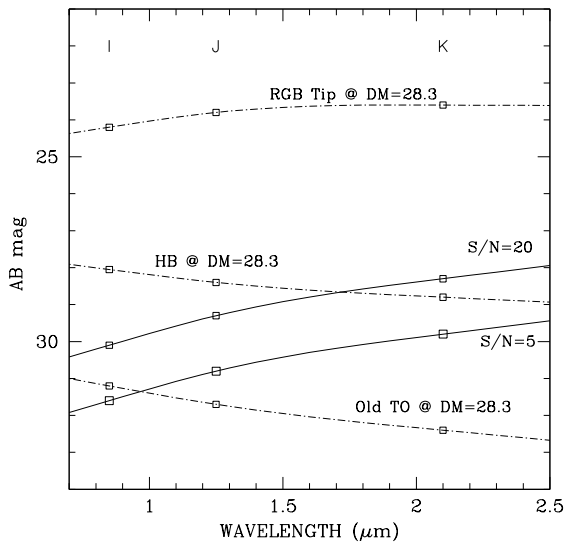}
\caption{Expected limiting magnitudes for point sources observed  with
  MICADO@E-ELT (solid lines) compared to the magnitude of RGB Tip, red 
  HB and 10 Gyr old MS TO stars at a distance of 4.6 Mpc
  (dot-dashed lines). The limiting magnitudes, relative to a S/N of 20
  (upper curve) and of 5 (lower curve), are computed for a total of
  5 hours integration time.} \label{limits}
\end{figure}

There  are  several  interesting  programs  concerning  the  study  of
resolved stellar  populations in external galaxies,  like the analysis
of multiple generations in Globular Clusters (see \citealt{piotto} and
references therein);  the determination  of the initial  mass function
and  its  variations  (e.g.  \citealt{kroupa}); the  recovery  of  the
spatially  resolved  SFH  in  isolated  and  in  interacting  galaxies
(e.g. \citealt{cioni}  for the Magellanic system);  tracing the galaxy
assembly process through the counts of resolved stars and detection of
streams (e.g.  \citep{annette}).  Here we focus on  the possibility of
deriving  global  information on  the  star  formation  history for  a
significative sample  of galaxies with a modest  time investment. This
means targeting distant galaxies, so  that the field of view samples a
fair  fraction of  their mass,  and  observing galaxies  in groups  or
clusters,  so  as to  enable  the  study  of a  representative  galaxy
population. On  these premises we  have concentrated on  the following
science cases:  A) deriving basic information on the  SFH in the  central region of  a disk
galaxy at  the  distance of the Centaurus  group; and B)
studying the metallicity distribution  at half of the effective radius
of an  elliptical galaxy  in the Virgo
Cluster.  For these cases  we have developed simulations of
stellar  photometry,  based   on  synthetic  stellar  populations  and
assuming  the expected  performances of  the Adaptive  Optics assisted
near-IR  camera  (MICADO,  \citealt{daviesmess})  for the  E-ELT  (see
details in Sect. 3.1).  The resulting CMDs were the analyzed
  to assess the impact of the photometric errors on the specific
  science goal.   

In section  2 we describe our  selected science cases.  Sect. 3 report
detail  of  our simulation  and  the  following photometric  analysis.
Results of  this study  are given in  Sect. 4 with  general conclusion
summarized in Sect. 5.

%\textbf{Figure CMD J-K of mamma mkm2.young with boxes superimposed}
\begin{figure}[t]
%\resizebox{\hsize}{!}{\includegraphics{cmd_young.eps}}
%\plotone{cmd_young.eps}
\plotone{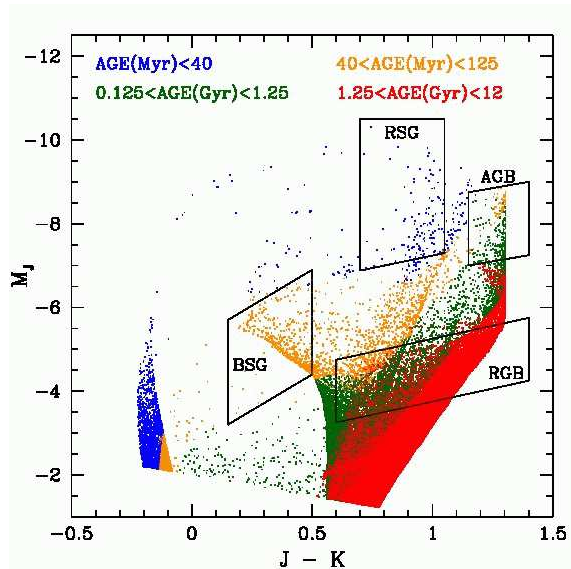}
\caption{Synthetic  CMD obtained  with a  constant rate of star formation over  the last
  12~Gyr,  a  Salpeter  initial  mass  function (IMF),  and  a  simple
  age-metallicity  relation which, starting  from $Z=0$,  goes through
  the  solar metallicity  ($Z_\odot$) at  4.5 Gyr  ago, and  reaches a
  value  of $Z=1.1  Z_\odot$ at  the current  epoch.   The simulation,
  computed with the YZVAR code  by G.P. Bertelli with the 2002 version
  of the Padova tracks \citep{girardi}, contains 200000 stars brighter
  than  $M_K = -2$,  and corresponds  to a 
%astrated 
 total mass of formed stars of $2.9\cdot 10^8$ M$_\odot$.  The four diagnostic boxes
 superimposed contain 122 (RSG), 598 (BSG), 340 (AGB) and 22198 (RGB)
 synthetic stars.}
\label{cmd_young}
\end{figure}

\section{The science cases}

%{\bf  show Luminosity Function and CMD }

The derivation of  the SFH in galaxies from the  CMD of their resolved
stars is  nowadays routinely performed  with the synthetic  CMD method
\citep{tosi,tolstoy,aparicio,zari,dolphin},  which aims  at minimizing
the   \textit{distance}   between   the   observed   stellar   density
distribution across  the CMD and  corresponding models by  varying the
strength  of  the  successive  star  formation  episodes.  The  method
basically  rests  upon age-dating  and  counting  stars in  diagnostic
regions on the CMD (see, e.g., \citealt{coimbra}; \citealt{book}), and
the  derived SFH  depends  on the  age  sensitivity of  characteristic
features sampled in the observed CMD \citep{carma}. Stellar age dating
is most accurate for stars at the main sequence (MS) turn off (TO); at
the other extreme, the red giant branch (RGB) collects stars born over
(almost)  the total  range of  ages present  in the  galaxy.  Also the
Horizontal Branch (HB)  is populated by stars in a  very wide range of
ages, but the  truly old populations can be  distinguished as RR Lyrae
stars or blue HB stars.

\begin{figure}[t]
%\resizebox{\hsize}{!}{\includegraphics{age_dist.eps}}
%\plotone{age_dist.eps}
\plotone{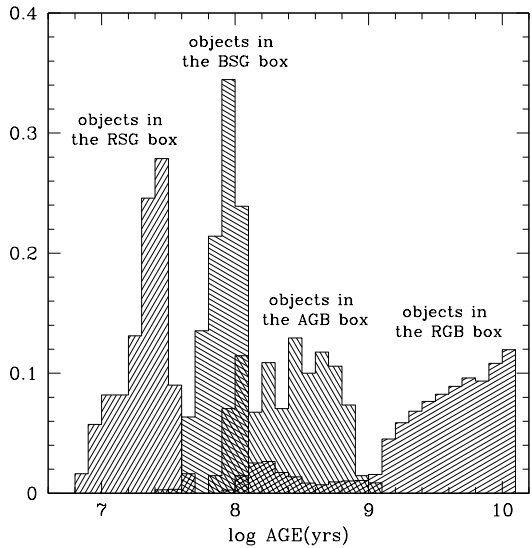}
\caption{Age distributions of the synthetic stars in Fig. \ref{cmd_young} 
which fall in the four diagnostic boxes, as labelled. Notice that each
distribution is normalized to the total number of objects in the
corresponding box.}
\label{age_dist}
\end{figure}

Fig. ~\ref{limits}  shows the limiting magnitudes  for isolated points
sources  that are  expected for  MICADO@E-ELT with  a 5  hours  of total
integration, together with
the typical magnitudes of RGB Tip  stars, red HB stars and MS TO stars
with an age of  10 Gyr, at a distance modulus of DM  = 28.3 
Comparing  the observational limits
with these  characteristic magnitudes we derive  the typical distances
up  to  which these  features  can be  detected  and  measured with  a
relatively  good S/N (i.e. $\sim 10$, see  Table
\ref{tab_mod}).  It appears  that we
could sample old MS TOs only within the Local Group and little beyond,
while for galaxies in the Centaurus  group (DM = 28.3) we will measure
HB  stars  with  high  S/N.  In the  nearest  galaxy  cluster  (Virgo,
DM=31.3),  the portion of  CMD accessible  with a  reasonable exposure
time is limited to the (upper) RGB and brighter magnitudes.  Actually,
HST already provides us CMDs down to these limits but only in sparsely
populated regions; for example in  Centaurus A, HB stars were detected
\cite{rejkuba05} with  $\sim 9$ hr  exposures, in a region  located at
$\sim  40$  Kpc from  the  center ($\mu_{\rm  K}  \sim$  26). At  this
distance, the portion  of the CMD at magnitudes  brighter than $M_{\rm
  I}  \sim -2$ has  been measured  and analyzed  for late  type dwarfs
around M83  \citep{grebel}.  Similarly, photometry down  to $\simeq$ 1
magnitude below the Tip of the RGB was obtained by \cite{caldwell} for
dwarf galaxies in Virgo.  The real gain we expect with MICADO concerns
the  ability of  deriving  data  of similar  quality  in high  surface
brightness regions  within galaxies, to the advantage  of sampling the
bulk of  the star  formation in medium  size and giant  galaxies.  The
simulations   described   in  this   paper   aim   at  checking   this
quantitatively for the two specific science cases described below.

\begin{table}[t]
\begin{center}
\caption{Maximum distance moduli (DM) to obtain accurate $J$ band photometry
  (S/N $\sim 10$) of isolated stars in various evolutionary stages, with 5 hours
of total integration with MICADO@E-ELT.\label{tab_mod}} 
\begin{tabular}{lr}
\tableline\tableline
Evolutionary Stage & DM\\
\tableline
Old MS Turn Off & 26.5 \\
RR Lyrae    & 27.5 \\
Red HB      & 30.0 \\
tip of RGB  & 34.5 \\
\tableline
\end{tabular}
\end{center}
\end{table}

%\textbf{Figure CMDs J-K , I-J of mamma mkm2.young with RGB box + Z distrib}
\begin{figure*}
%\resizebox{\hsize}{!}{\includegraphics{cmd_old.eps}}
\resizebox{\hsize}{!}{\includegraphics{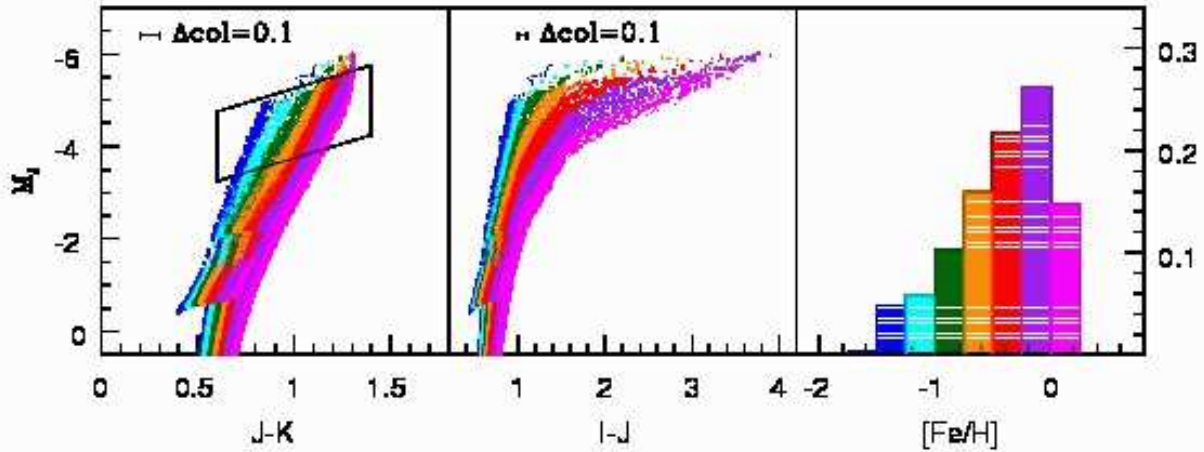}}
\caption{Synthetic  CMDs for  a  stellar population  with  a flat  age
  distribution  between  10  and   12  Gyr,  and  a  wide  metallicity
  distribution \citep{rejkuba05} plotted in the right panel.  Dots are
  coloured according  to their metallicity, with the  same encoding as
  in the right panel.  The simulation, computed with the YZVAR code by
  G.P. Bertelli  with the 2002  Padova tracks database and  adopting a
  Salpeter IMF,  contains 200000 stars  brighter than $M_K =  +1$, and
  corresponds  to  a total  mass  of formed stars of  6.8  $\cdot  10^7$
  M$_\odot$.  The  RGB  diagnostic  box  on  Fig.  \ref{cmd_young}  is
  reproduced on  the left  panel. Notice that  the color range  of the
  infrared  CMD  (left  panel)  is  much narrower  than  that  of  the
  optical-infrared CMD (central panel).}
\label{cmd_old}
\end{figure*}

\subsection{The SFH in a disk galaxy in the Centaurus group}

Fig.~\ref{cmd_young} shows a synthetic CMD which can be considered
as representative of the stellar population in the disk of a late type
galaxy, with 
the  synthetic   stars  color  coded   according  to  their   age,  as
labelled. 
We  examine here the  possibility of recovering 
basic information on the  SFH for
such a stellar population placed at  a distance of $\simeq$ 4.5 Mpc.
Four diagnostic
boxes (see  Table \ref{tab_boxes})  are superimposed on  the synthetic
CMD,  with the aim  of targeting  specific age  ranges, i.e.   the Red
Supergiant (RSG) box for  the youngest component; the Blue Supergiants
(BSG) and Asymptotic Giant Branch (AGB) boxes for the intermediate age
components; the RGB box for the old component. The age
distributions of
the  stars  falling   in  each  of  the  four   boxes  is  plotted  on
Fig. ~\ref{age_dist}, showing that there  is little age overlap in the
various boxes. The simulation allows us to estimate the average number
of stars per unit stellar mass formed in the age range sampled by each
box, hereafter  referred to as specific  productions (\Pbox), reported
in Table \ref{tab_boxes}.
Star counts  in these diagnostic  boxes performed on  an observational
CMD allow us to recover  quantitative information of the SFH in
the  observed field  over the  whole  galaxy lifetime:  it suffice  to
divide  the star  counts by  the appropriate  specific  productions to
derive the  average mass  turned into stars  in the  corresponding age
range. This method is not as sophisticated as the full synthetic CMD
method, which fits the detailed stellar distribution using small boxes, but it's much more straightforward and easily applicable to
large datasets. At any rate,
the  reliability of the  derived SFH depends on the  completeness of
the counts within the boxes,  sensitive to the photometric error which
\textit{moves}  the  stars  from  or  into  adjacent  regions  of  the
CMD. Therefore  we expect a dependence  of the result  on the crowding
conditions,  i.e. on  the  surface brightness  of  the sampled  galaxy
region. In the following we test 
how the star counts in these boxes are affected
for a stellar field
in the  very center of  a disk galaxy  located in the  Centaurus Group
adopting a distance modulus on $m-M=28.3$ and no extinction.

\begin{table}
\caption{Properties of the diagnostic boxes for the \textit{YOUNG}
  population science case: column (2) reports the age  range sampled, column (3)
  the   average  specific   production   (for  a   Salpeter  IMF)   in
  $M_\odot^{-1}$, columns  (4) and (5):  number of synthetic  stars in
  the input and  output CMD of our simulation, which refers to
  a total stellar mass of $\sim$ 2.6 $\times 10^{7}$ \msun\ in the FoV.}
\begin{tabular}{lcccc}
\tableline\tableline
Box & Age Range & $\Pbox$ & N$_{\rm inp}$ & N$_{\rm out}$ \\
\tableline
RSG & $\leq$ 40 Myr & 1.26 $\times 10^{-4}$ &14 & 14\\
BSG & ( 40 to 120) Myr & 2.91 $\times 10^{-4}$ & 59 & 59\\
AGB & ( 0.12 to 1) Gyr & 1.25 $\times 10^{-5}$ & 35 & 35\\
RGB & $\geq$ 1 Gyr & 8.54 $\times 10^{-5}$ & 1933 & 1929\\
\tableline\tableline
\label{tab_boxes}
\end{tabular}
\end{table}

\subsection{The metallicity distribution for stars in a giant elliptical in Virgo}

Fig.~\ref{cmd_old} shows  the luminous  portion of the  CMD of  an old
stellar  population  with a  wide  metallicity  distribution, akin  to
elliptical galaxies.
Of the  four diagnostic boxes  drawn on Fig.~\ref{cmd_young}  only the
RGB box  is populated,  and includes 3232  stars, yielding  a specific
production of  $P_{\rm RGB} =4.75 \, 10^{-5}  M_{\odot}^{-1}$ for this
stellar population.
The absence of stars in the RSG, BSG and AGB boxes implies (virtually)
no  star formation  at ages  younger  than $\simeq$  1.5 Gyr,  whereas
dividing the star  counts in the RGB box by  $P_{\rm RGB}$ one obtains
the mass transformed into stars in the field.
More interestingly, the color distribution of the bright RGB stars can
be  used  to recover  their  metallicity  distribution,  as in,  e.g.,
\cite{harris}.  The different metallicity bins separate much better on
the central panel of Fig. \ref{cmd_old}, due to the higher temperature
sensitivity  of   the  wider  color  baseline.  In   this  respect  an
optical-infrared   color   provides   the   best  leverage   for   the
determination of  the metallicity distribution  of the RGB  stars. The
horizontal bars in  the left and central panels  of Fig. \ref{cmd_old}
show the size of a 0.1  mag error on the respective color: clearly the
photometric accuracy needed to recover the metallicity distribution is
less demanding in  the optical-infrared CMD than in  the infrared CMD.
At any rate, photometric errors  smear out the true color distribution
of  the stars,  thereby  affecting the  derivation  of the  underlying
metallicity  distribution.  Thus, the  accuracy  of this  distribution
determined  from the  analysis of  the  CMD is  sensitive to  crowding
conditions. In  the following  we test the  effect of  the photometric
errors for a  stellar population located at half  the effective radius
of  an elliptical  galaxy in  the Virgo  Cluster, adopting  a distance
modulus of $m-M=31.3$ and no extinction.

\begin{figure}[t]
\begin{center}
%\plotone{maory_psf.ps}
\plotone{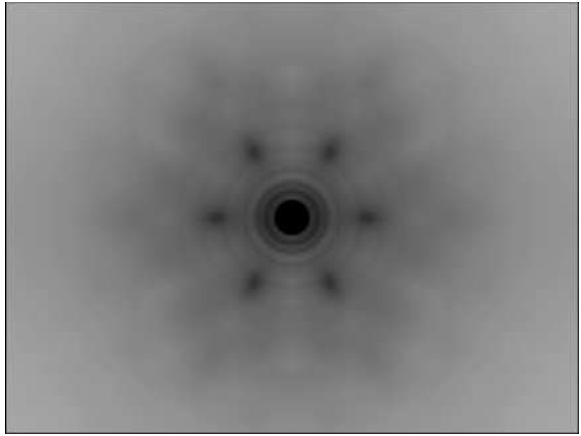}
\caption{MAORY PSF in J band (FoV 0.05 arcsec)  MICADO@E-ELT.
\label{fig_psf}}
\end{center}
\end{figure}
\begin{figure}[h]
\begin{center}
%\plotone{ee_maory.eps}
\plotone{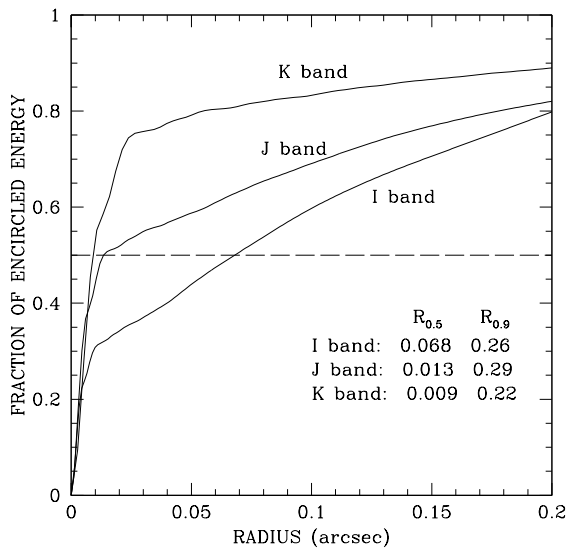}
\caption{Radial growth of the encircled energy of the PSFs adopted for
  our synthetic frames. The inset shows the radii (in
  arcsec) which include 50 \% (R$_{0.5}$) and 90 \% (R$_{0.9}$) of the total energy.
  \label{ee_maory}}
\end{center}
\end{figure}

\section{Simulations and Analysis}

Simulated  images were constructed  with a  tool\footnote{The Advanced
  Exposure Time Calculator, AETC; http://aetc.oapd.inaf.it/} developed
during  the Phase  A study  of MICADO  \citep{ffu}.
The simulator  is able to  generate synthetic
frames with a large variety  of telescope and instrument parameters as
well as for input field of stars and/or galaxies.  In the MICADO@E-ELT
configuration all the relevant parameters  of the telescope and of the
instrument   are  included,  as   well  as   the  conditions   of  the
observations.  For  the simulated images  presented here we  assumed a
1277  m$^{2}$ collecting  area, a  total  throughput of  0.4 and  0.39
respectively in  the $I$ and $K$  bands, a read noise  of 5 electrons,
and a plate-scale of 3 mas per pixel.  We assumed the typical near- IR
sky  background at Paranal  (Chile) and  included the  contribution of
thermal  emission in  the near-  IR bands;  specifically we  adopted a
background os 20.1,  16.3 and 12.8 mags in the $I$,  $J$ and $K$ bands
respectively.

\begin{figure}[t]
%\begin{center}
%\plotone{image_G.eps}
\plotone{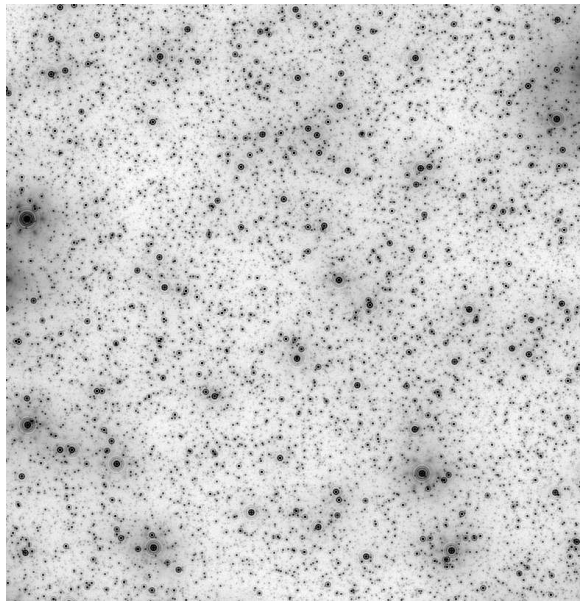}
\caption{ Central region ($1.5 \times 1.5$ arcsec) of the simulated image of
 the disk galaxy (case A) in the $J$ band as would be observed in 5 hr exposure using
 MICADO@E-ELT.  The surface brightness is $\mu_{J} $= 18.9 $\Box^{\prime\prime}$.}
\label{caseg_jframe}
%\end{center}
\end{figure}

\begin{figure}[h]
%\begin{center}
%\plotone{image_E.eps}
\plotone{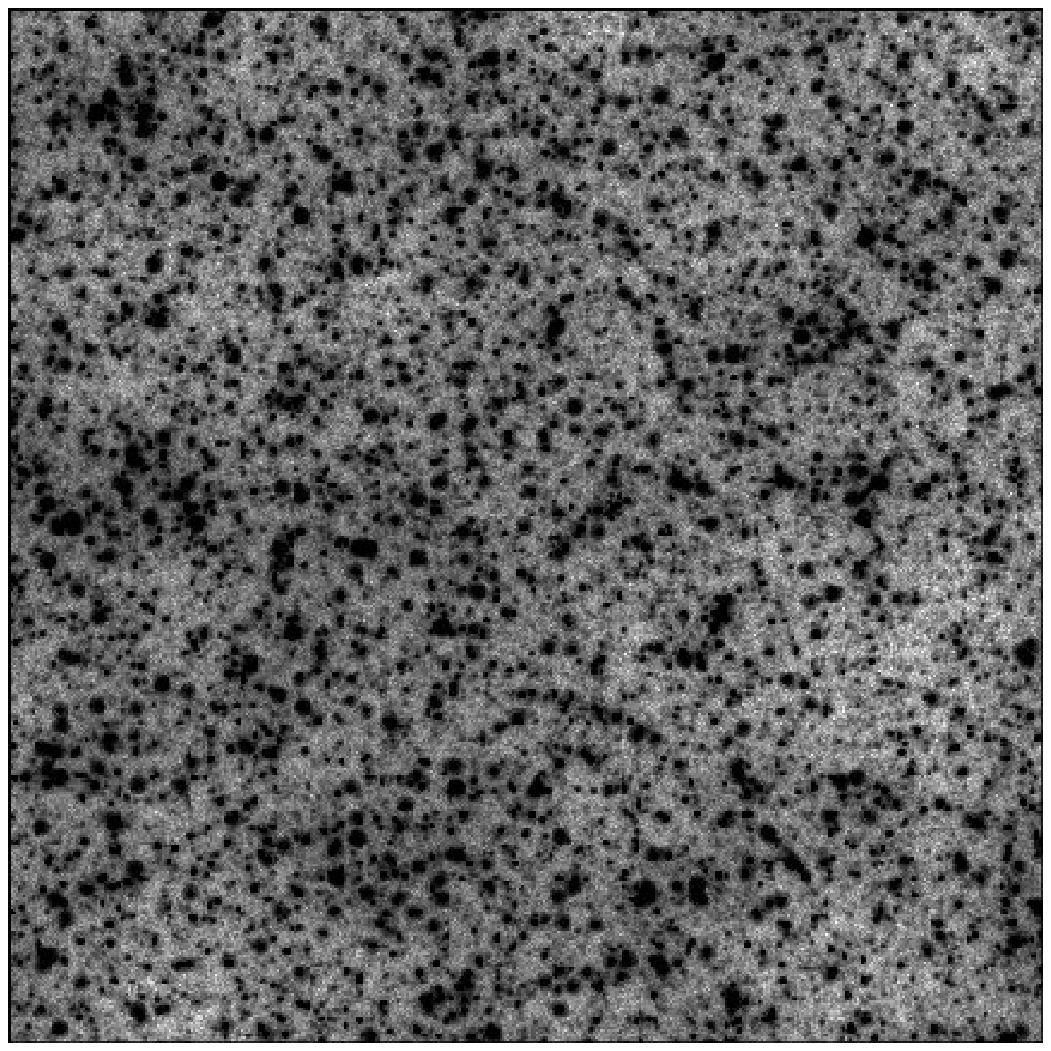}
\caption{ Central region ($1.5 \times 1.5$ arcsec) of the simulated image of
 the elliptical galaxy (case B) in the $J$ band as would be observed in 5 hr exposure using
 MICADO@E-ELT. The surface brightness is $\mu_{J}$ = 18.8  $\Box^{\prime\prime}$.}
\label{casee_jframe}
%\end{center}
\end{figure}

\subsection{MICADO}

MICADO is the Multi-AO Imaging Camera for Deep Observations, designed to work
with adaptive optics (AO) on the 42m E-ELT \citep{davies}. 
MICADO is  optimized for  imaging at the  diffraction limit,  and will
fully sample  the 6-10mas  full width half  maximum (FWHM) in  the J-K
bands. With a throughput exceeding  60\% its sensitivity at 1-2 $\mu$m
will be  comparable to that of  JWST (James Webb  Space Telescope) for
isolated point  sources, but MICADO's exquisite  resolution will allow
us to  investigate regions  with a much  higher crowding,  compared to
JWST.  The instrument is able  to image a relatively wide ($53 \arcsec
\times 53  \arcsec$) Field  of View,  and a large  number of  wide and
narrow band filters is foreseen, ranging from the $I$ to the $K$ band.

The simulations  assume that the  images are obtained using  MICADO in
conjunction  with  MAORY,   the  post-focal  adaptive  optics  modules
currently under study for the European Extremely Large Telescope.  The
MAORY \citep{diolaiti, foppiani} module  uses six laser guide stars to
produce  multi-conjugate  AO (MCAO)  correction.  We  assumed the  PSF
computed  with  a seeing  of  0.6 arcsec \footnote{
http://www.bo.astro.it/maory/Maory/Welcome.html}, with FWHM of 4 mas, 5
mas and  9 mas in the $I$,  $J$ and $K$ bands  respectively. Since the
Field of  View of our simulation  is small compared  with the expected
PSF variation  on the total  field of MICADO,  we assumed a  fixed PSF
over the  whole simulated  frame. Specifically, we used the MAORY central
  PSF mapped over 2048 $\times$ 2048 pixels. In Figure  \ref{fig_psf} we  show a
representative image of the adopted PSF in J band.  The PSF in the $I$
and  $K$ bands  have a  very similar  shape but  different  Strehl
(i.e. the ratio of peak diffraction intensities of an aberrated versus perfect wavefront)
and encircled energy  values. Fig. \ref{ee_maory} shows  the radial growth
of  the  encircled  energy  in  the photometric  bands  used  for  our
simulations. These  distributions present  a very sharp  central peak,
which becomes progressively more pronounced  going from the $I$ to the
$K$ band,  reflecting the trend  of the Strehl ratio.   A quantitative
characterization of the PSF is included in Fig.  \ref{ee_maory}.

\begin{table*}
\caption{Parameters of the simulated stellar populations. Column (2): ratio
  between the total mass of formed stars and the $B$-band light in solar units;
  columns (3) to (6): integrated colors; columns (7) to (9): adopted
  surface brightness, distance modulus and Field of View; column (10):
  total mass (in solar units) of the simulated stellar population following from
  the previous entries.}
\begin{tabular}{lccccccccc}
\tableline\tableline
Stellar Population & $\msf/L_{\rm B}$ & $B-V$ & $B-I$ & $B-J$ & $B-K$
& $\mu_{\rm B}$ & DM & FoV & \msf \\
\tableline
$YOUNG$ & 1.52 & 0.48 & 1.29 & 2.16 & 2.99 & 21.07 & 28.3 & 12''
$\times$ 12'' & 2.66 $\times 10^7$\\
$OLD$  & 7.05 & 0.88 & 1.97 & 2.84 & 3.66 & 21.6 & 31.3 & 3'' $\times$
3'' & 7.53 $\times 10^7$ \\
\tableline\tableline
\label{tab_csp}
\end{tabular}
\end{table*}

\subsection{Input stellar populations}

Table  \ref{tab_csp}  lists  the   properties  of  the  input  stellar
populations of the two science cases:  A) the central region of a disk
galaxy  in  the  Centaurus  group  (hereafter  \textit{YOUNG}  stellar
population) and B) a region located at half of the effective radius of
an  Elliptical in  the Virgo  cluster (hereafter  \textit{OLD} stellar
population). Illustration of CMDs  for the two stellar populations are
shown  in  Figs~\ref{cmd_young}  and  ~\ref{cmd_old},  where  we  also
specify the adopted star formation histories.

The  input stellar  lists are  generated from  theoretical simulations
following a  procedure described  in detail in  the Appendix;  here we
only  mention that  special care  was devoted  to ensure  (i) adequate
sampling of the short lived evolutionary stages, and (ii) completeness
of the  input stellar list  in all the  photometric bands used  in the
specific science  case. The size  of the simulated  population follows
from the proportionality  between the number of stars  brighter than a
given  magnitude  limit  and  the  total  luminosity  of  the  stellar
population sampled by the synthetic frame:

\begin{equation}
L_{\rm B} = FoV \, 10^{-0.4*(\mu_{\rm B} -5.48 - {\rm DM)}} \,\,\, 
L_{\rm B,\odot}
\label{eq_lb}
\end{equation}

\par\noindent where $FoV$ is the  Field of View in square arc seconds,
$\mu_{\rm B}$ is  the (un reddened) surface brightness  of the stellar
population, DM  is the distance modulus,  and we have  adopted a solar
absolute magnitude in the B band of $M_{\rm B,\odot} = 5.48$.

In order to keep the input  stellar list within a manageable size only
objects brighter  than a threshold  magnitude are considered  as point
sources,  while  the remaining  light  of  the  stellar population  is
distributed over the frame as  a pedestal, with its associated Poisson
noise. The threshold  magnitude is chosen $\sim$ 1.5  mag fainter than
the  limiting  magnitude (  S/N=5  )  of  the telescope  +  instrument
combination  for the  assumed  exposure time.  This  ensures that  the
effect of  blending of  stellar images is  well described also  at the
faint end of the luminosity function.

\begin{figure}[t]
%\plotone{lf_g.eps}
\plotone{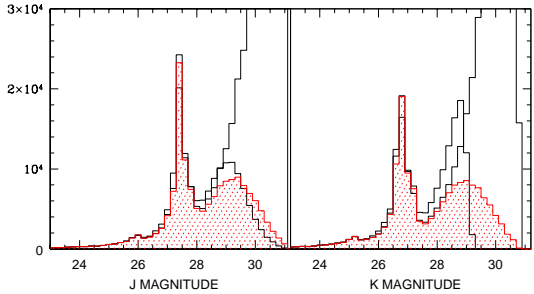}
\caption{Input and  output luminosity functions in the  $J$ (left) and
  $K$ (right) bands for  the \textit{YOUNG} stellar population science
  case. The thin lines show the luminosity function of the whole input
  list.  The shaded  histogram highlights  the luminosity  function of
  those input stars which have  been position matched in the reduction
  process.   The  thick lines  show  the  output luminosity  function.
  Notice the effect of blending particularly evident at the faint end,
  so that, e.g., stars with input  magnitude as faint as $K \simeq 30$
  are  detected  in  the  reduction  process, but  are  measured  much
  brighter than they are.}
\label{lf_g}
\end{figure}

\subsection{Synthetic Frames}

The  AETC tool  with the  MICADO  configuration was  used to  generate
images in  the $I$,  $J$ and $K$  bands for  the two science  cases in
Table  \ref{tab_csp}. Notice  that the  mass  to light  ratio and  the
colors are univocally determined by the adopted SFH, while the stellar
mass  (\msf), sampled  by the  frame,  depends on  the chosen  surface
brightness, distance and $FoV$. For the \textit{YOUNG} case, the input
stellar list consists of 644483  stars brighter than $K$=30.7; for the
\textit{OLD} case,  it counts 125058  stars brighter than  $K$=31. The
input  lists specify mass,  age, metallicity,  magnitudes in  the $I$,
$J$,  and   $K$  bands,  and   coordinates  on  the  frame   for  each
star. Synthetic  frames are generated  from them in  the corresponding
bands  and  Figures  \ref{caseg_jframe}  and  \ref{casee_jframe}  show
sections of the $J$ band simulated images. The wide magnitude range of
the objects in Fig.  \ref{caseg_jframe} is readily visible, with a few
very bright  stars showing the  characteristic pattern of the  PSF. It
can be  noticed that many objects  are detected under the  halo of the
bright  stars  due  to the  narrow  core  of  the  PSF. The  image  in
Fig.  \ref{casee_jframe} has  a very  different appearance,  with many
stars of comparable brightness: this reflects the different properties
of the luminosity function of the two populations.

\begin{figure*}[t]
%\resizebox{\hsize}{!}{\includegraphics{dmag_g.eps}}
\resizebox{\hsize}{!}{\includegraphics{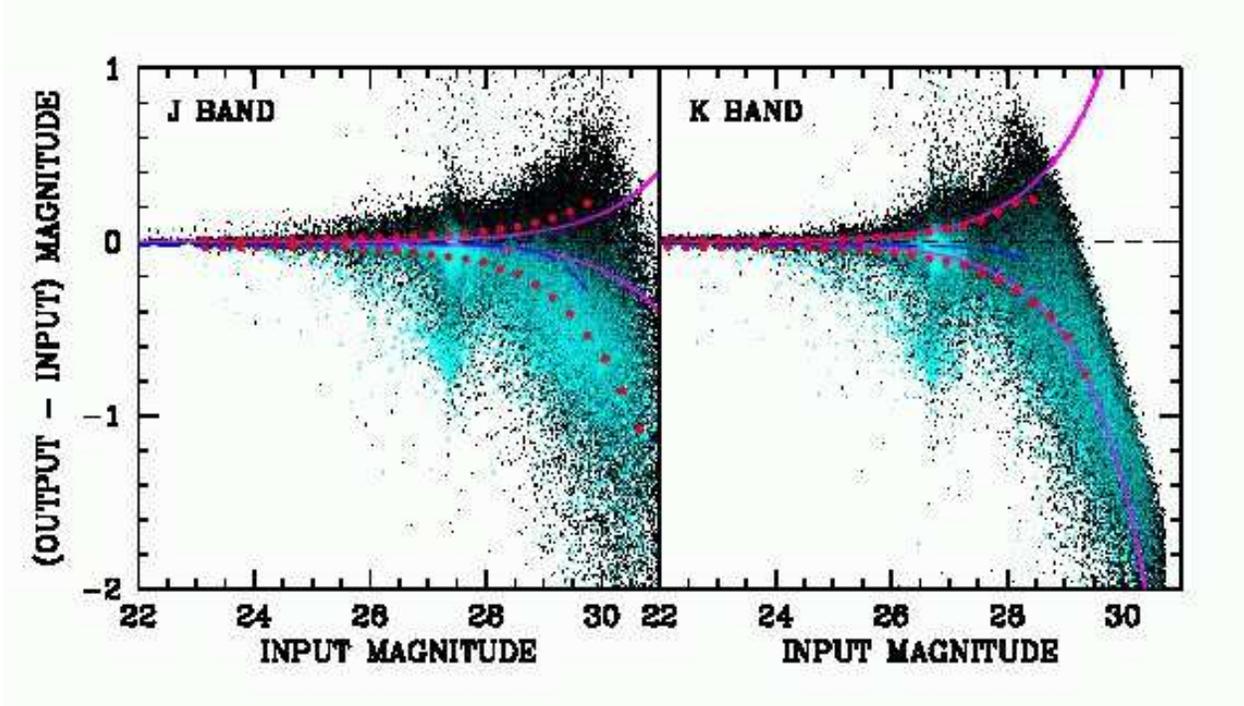}}
\caption{Photometric error  as function of the input  magnitude in the
  $J$  (left) and $K$  (right) bands  of all  matched sources  for the
  \textit{YOUNG}  stellar population  case.  Dark  (light)  points are
  sources with a single  (multiple) candidate counterpart in the input
  list.   Solid  lines show  the  1$\sigma$  uncertainty for  isolated
  stars.  Dashed  lines show  the median error  of the  matched stars:
  notice  that because of  blending more  than half  of the  stars are
  measured  brighter than  they are  at $J,  K \gsim  28$.  The filled
  circles  show  the 1$\sigma$  widths  of  the  negative and  of  the
  positive  error distributions.  The median  error and  the +$\sigma$
  width are plotted only up to  the value of input magnitude for which
  the error distribution on the positive side is fully sampled.}
\label{dmag_g}
\end{figure*}

\subsection{Photometric Analysis and Methodology}

We performed  the extraction  of the PSF  photometry on  the simulated frames
using  the  \cite{Stetson94}  suite  of  programs:  DAOPHOT,  ALLSTAR,
ALLFRAME,  and  related  supporting  software.   The  usual  steps  of
detecting  the sources  in the  images, creating  the PSF  and finally
measuring  the  PSF  photometry  were executed  via  a  semi-automated
procedure. To avoid  the introduction of any possible  bias, the image
simulations were performed by a person different from the one
who carried on the photometry  of the fields. Therefore  the analysis
was performed with no  previous knowledge of the  simulated images.
Only the "observational" parameters  present in the
header  of the  image has  been used,  like the  number of  summed and
averaged exposures, the gain of the detectors, the read out noise etc.

Performing PSF  photometry with DAOPHOT  requires two main  steps: the
construction  of  the  PSF  and  the  extraction  of  the  photometric
measurements. For the PSF creation we  used 500 stars per image and we
calculated     all    the    6     different    options     for    the
analytical\footnote{DAOPHOT  allows  the  calculation of  6  different
analytical  model  PSF  via  its {\tt  analytic  function}  parameter:
Gaussian,  Moffat  1.5, Moffat  2.5,  Lorentz,  Penny  4 and  Penny  5
functions.}  modeling of  the PSF, letting the program  decide for the
best PSF model  to be used based on the  overall minimum $\chi^2$. 
The procedure was iterated up to 5 times. At each iteration we
instructed DAOPHOT to subtract all the PSF neighboring stars from the
original image and used this new image to extract the updated PSF
model. We also discarded all PSF candidates with bright neighbors,
or with saturated/dead pixels. This procedure turned out 
very effective, typically converging to a suitable PSF within 3
iterations. The particular shape
of the MAORY  PSF with its sharp core and extended  wings forced us to
use a DAOPHOT PSF dimension much  larger than the one used for typical
gaussian-like PSF.   For the latter case  the PSF radius  of DAOPHOT is
$4\div5\times$ the  FWHM of the stars,  while in the case  of MAORY we
had to use a  factor of $15\times$ the FWHM of the  core.  In this way
we  enclose  $\simeq  95\%$  of   the  stellar  flux  within  the  PSF
radius. The analytic part of the of PSF was reduced to a minimum allowing for
the DAOPHOT fitting radius a value of $\simeq0.9\times$ that of the core FWHM.
The iterative procedure outlined above ensures an accurate
description of the PSF, including its wide wings.
The FWHM  of the  PSFs finally  extracted by  DAOPHOT  are of
$\simeq$ 1.8, 2.2  and 3.8 pixels in the $I$, $J$  and $K$ band images
respectively,  with  a  very  small  difference  in  the  two  science
cases. These measured FWHM are  only slightly larger than those of the
input PSFs (see Sect. 3.1). Although we set a fixed (non position
dependent) PSF, we allowed for internal variability of the
reconstructed DAOPHOT PSF.

The  final  extraction  of   the  photometry  is  performed  with  the
ALLSTAR+ALLFRAME programs.  ALLSTAR extract  the PSF magnitudes on the
image using the model PSF reconstructed by DAOPHOT. When more than one
image of the  same field is available (even  in different filters) the
extracted photometry can be used as input to ALLFRAME, which repeats
the ALLSTAR procedure simultaneously on all images.  This considerably
improves  the final  photometric precision  because of  the  much more
significant  statistical  treatment  of  each  detected  star  in  the
observed field, both position  and photometry are being constrained by
all images.  The  use of ALLFRAME allowed us to extend  the final list of
the  measured  objects  at a  $\sim$  15  \%  fainter limit  than  the
detection threshold in a single band.

The previously described procedure  allowed the measurement of the PSF
photometry of each image in an automatic, un-supervised mode requiring
only  one input parameter:  the extraction  threshold. We  usually set
this parameter to $3.0\times\sigma$ of the background, while all other
parameters were set internal to the procedure. The final product is a
catalogue of  magnitudes measured with  ALLFRAME operating on  the $J$
and $K$ images for the \textit{YOUNG} stellar population science case,
on the  $I$, $J$ and  $K$ band images  for the \textit{OLD}  one.

\begin{figure}[t]
%\plotone{lf_e.eps}
\plotone{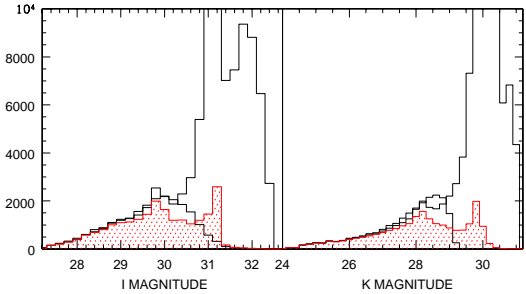}
\caption{Input and  output luminosity functions in the  $I$ (left) and
  $K$ (right)  bands for  the \textit{OLD} stellar  population science
  case.  Thin and  thick lines  show  respectively the  input and  the
  output  luminosity functions;  the shaded  histogram  highlights the
  luminosity function  of those input  stars which have  been position
  matched. Notice that the large distance prevents us to recognize the
  populous HB/clump feature.}
\label{lf_e}
\end{figure}

\section{Results}

\subsection{Photometric Quality}

Having  full knowledge  of the  input star  list we  can  evaluate the
quality  of  the  photometry  object  by  object.  In  order  to  pair
detections  to input  sources it  is  necessary to  define a  matching
algorithm, a  task which  is not completely  trivial. We use  a search
radius of 1  pixel to select candidate mates  of each detected source;
within  such area  we may  find  (a) no  input source,  (b) one  input
source,  (c) more  than  one  input source.  Case  (a) corresponds  to
spurious detections,  due to, e.g.,  noise spikes, or  interference of
the wings of the PSF of neighboring bright stars; case (b) corresponds
to a one to one match, while  in case (c) we have the embarrassment of
more than  one possible (input star)  mate. We adopt  the criterion of
choosing as counterpart the brightest among the candidates.
Different criteria  could be envisaged  (e.g. imposing a limit  on the
difference  between  the input  and  output  magnitudes), all  however
getting  into  trouble  at  faint  magnitudes,  when  approaching  the
confusion limit. Our choice has the merit of being totally independent
of the magnitude difference between the input and output source, which
is in fact the subject of the analysis described in the following.

\subsubsection{The Young Stellar Population at 4.6 Mpc}

The simulated images for this science case include $\sim$ 650000 stars
with $J, K \leq 31$;  the photometric measurement yields $\sim$ 163000
sources detected in $J$ and $K$  of which $\sim$ 159000 have a matched
counterpart in the input list, while  $\sim$ 4000 (i.e. $\sim$ 2 \% of
the  detections) are  spurious  objects. Most  of these  \textit{fake}
sources are located around input bright stars, and are features of the
PSF (see Fig. \ref{fig_psf})  misinterpreted as stars by the reduction
package   because    of   a    non   perfect   subtraction    of   the
PSF. Fig.  \ref{lf_g} compares the  luminosity function of  the output
catalogue (thick  lines) to  that of the  input data (thin  line). The
luminosity function appears very  well recovered brighter than $J \sim
26.5, K \sim  25.5$, while a strong incompleteness  is evident fainter
than $J,  K \sim 29$. In  the intermediate magnitude  range the output
luminosity function  counts an excess  of stars compared to  the input
data because of  blending, i.e. a luminosity increase  of the detected
source due to overlapping PSFs, as argued below. In Fig. \ref{lf_g} we
also show the  luminosity function of the input  stars which have been
position matched  (shaded histogram): clearly  only a fraction  of the
input stars are recovered in each bin, which is the true completeness,
or the probability of detecting  a star of given input magnitude.  Due
to  photometric errors,  the position-matched  stars can  be  found in
different bins, typically brighter than their true value, which is the
reason why  the output  luminosity function exceeds  the input  one at
most  magnitudes except for  the brightest  bins, where  photometry is
very accurate. At the faintest magnitudes incompleteness is boosted by
the low S/N and stellar background, and the output luminosity function
drops to zero.

The  photometric quality is  further illustrated  in Fig.~\ref{dmag_g}
where we compare the magnitudes  of the matched objects. The brightest
stars  are clearly  recovered with  a small  error, but  as  the input
magnitude increases,  the distribution of the  discrepancies widens as
the stars are measured either fainter or brighter than they are. Close
to  the faint  end  of  the input  magnitude  distribution the  source
detection is favored  if the local background is  low, and/or the star
is  blended  with other  sources.  Therefore,  at  the faintest  input
magnitudes,  positive  errors  (i.e.  output magnitude  $\gsim$  input
magnitude) become  under-sampled, and  below some threshold,  stars are
recovered  \textit{only}  if  measured  brighter than  they  are.  The
slanted limit  in the right panel of  Fig.~\ref{dmag_g} corresponds to
the detection threshold  used in the reduction of  the $K$ band image,
i.e. an  output magnitude  of $\simeq$ 29.  The effect of  blending is
present  over a  wide magnitude  range, and  leads to  an asymmetrical
distribution  of the photometric  errors, which  extends over  a wider
range on the  negative side compared to the  positive side. Notice for
example that the clump stars (at  $J \simeq 27.3$) can be recovered up
to 1  mag brighter, or up to  0.5 mag fainter than  they are. Although
the average error  at this magnitude level is very  close to zero, and
the peak of the luminosity  function due to clump stars well recovered
by the photometry (see Fig.  \ref{lf_g}), the spurious excess of stars
above the  true clump  can be misinterpreted  when analyzing  the CMD,
e.g. as  due to a younger stellar  generation.  Fig.~\ref{dmag_g} also
shows  that negative  errors are  much more  frequent for  sources for
which multiple candidate counterparts where found in the input stellar
list,  again demonstrating  the constructive  interference  of stellar
crowding,  which leads  to  an  artificial  brightening  of  the  detected
sources.

\begin{figure*}[t]
%\resizebox{\hsize}{!}{\includegraphics{dmag_e.eps}}
\resizebox{\hsize}{!}{\includegraphics{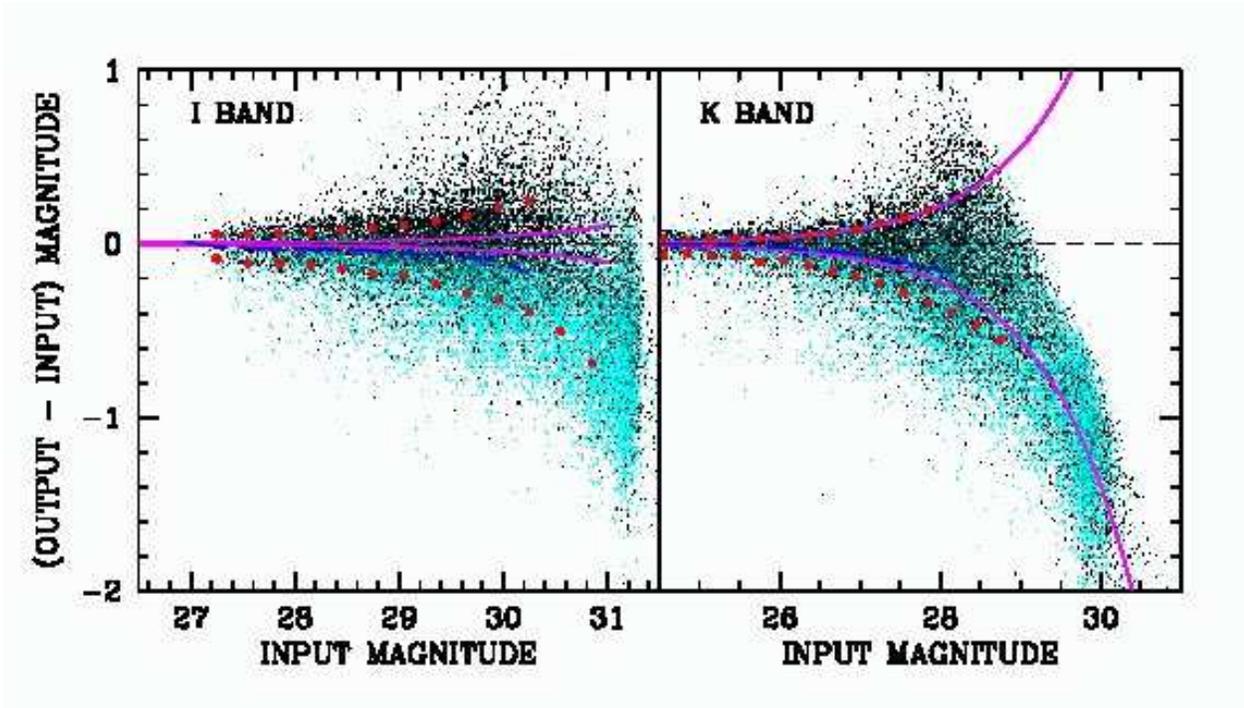}}
\caption{Photometric errors  as function of  the input $I$  (left) and
  $K$ (right)  magnitude, for sources with single  match (dark points)
  and with multiple match (light points) on the image. The dashed line
  shows the  median error; the big  dots show the  1$\sigma$ widths of
  the  error  distributions   computed  separately  for  positive  and
  negative errors.   Notice that clump stars with  input magnitudes $I
  \sim 31$,  $K \sim  30$ are identified  only if their  luminosity is
  boosted by blending, i.e. only with large and negative errors.}
\label{dmag_e}
\end{figure*}

The asymmetry  of the  photometric errors is  traced by the  median of
their distribution  (which becomes progressively more  negative as the
input magnitude gets  fainter), as well as by  the 1$\sigma$ widths of
the error distributions, computed independently for the stars measured
brighter (lower  locus in Figs \ref{dmag_g} and \ref{dmag_e}) and those measured fainter  (upper locus) than
they truly are. The asymmetry of the 1$\sigma$ width appears much more
pronounced  in the  $J$ rather  than in  the $K$  band because  of the
higher  background in  the  latter filter.  Actually  the total  error
mainly results from  the shot noise of the  sky and thermal background
plus the error associated to the underlying stellar population. In the
$K$ band  the first contribution  (which induces a  symmetrical error)
largely dominates over the second. The opposite holds in the $J$ band,
where the  effect of the  underlying stellar population can  be better
appreciated as an asymmetrical error distribution due to blending.

\subsubsection{The Old Stellar Population at 18 Mpc}

The simulated images for this science case include $\sim$ 125000 stars
with  $K \leq  31$  and $J$,  $I$  magnitudes down  to  31.7 and  32.5
respectively. The photometric  measurement yields $\sim$ 22350 sources
detected in  $I$, $J$  and $K$  of which $\sim$  21700 have  a matched
counterpart in the  input list, while $\sim$ 650 (i.e.  $\sim$ 3 \% of
the detections) are  spurious objects. Also in this  case, most of the
spurious detections come from a  non perfect PSF subtraction, and also
in this  case the spurious sources  are a small fraction  of the total
number of  objects, so that the  overall appearance of the  CMD and of
the LF is not significantly affected.

As  for  the \textit{YOUNG}  stellar  population, Figs~\ref{lf_e}  and
~\ref{dmag_e} illustrate the  photometric quality for the \textit{OLD}
stellar population. This science case differs from the previous one in
the distance modulus (3 magnitudes fainter) and in the magnitude range
covered by  the input star list,  because of the absence  of the young
and intermediate  age components. The input  luminosity function (thin
lines in Fig.  ~\ref{lf_e}) does include the HB, at  $31 \lsim I \lsim
32$, $29.5  \lsim K \lsim  30.5$, but only  very few of its  stars are
position matched  (see shaded histogram),  and in most cases  they are
assigned  a  too bright  magnitude.  The  migration  of stars  towards
brighter bins along the luminosity function due to blending is evident
also in  this case, and can  be appreciated as  the difference between
the  shaded region  and  the output  luminosity  function (thick  line
histogram). It is interesting to note that the cut off of the $I$ band
output luminosity  function is more smooth  than that of  the $K$ band
one. This  is due to the  measurement procedure adopted  which, once a
source  is detected  on the  $K$ image,  forces the  detection  at the
shorter   wavelengths,  so  that   the  adopted   detection  threshold
translates  into a  sharp  cut off  only  in the  $K$ band  luminosity
function.

\begin{figure}[t]
%\plotone{cmd_g.eps}
\plotone{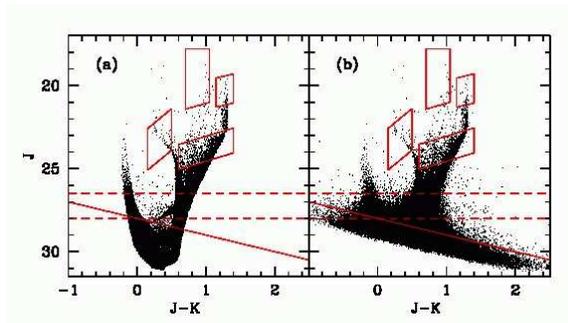}
\caption{Input (a) and output  (b) CMDs for the \textit{YOUNG} stellar
  population science case.  The four diagnostic boxes are  the same as
  in Fig. \ref{cmd_young}.  The  $J=26.5,28$ (dashed lines) and $K=28$
  (solid lines) loci are drawn  to better appreciate the effect of the
  photometric error on the CMD features.}
\label{cmd_g}
\end{figure}

The  distribution  of  the  photometric  errors  (Fig.  ~\ref{dmag_e})
presents   similar  characteristics   to  those   described   for  the
\textit{YOUNG} stellar  population case:  due to blending,  the median
error is  negative almost all  along the magnitude range  covered, and
the  error  distributions  are   skewed  towards  the  negative  side.
Compared  to  the  \textit{YOUNG}  one,  the  \textit{OLD}  population
science case is  characterized by a much higher  crowding, in spite of
the similar surface  brightness, because of the $\sim$  4 times larger
distance.  For this reason, the  error distributions are wider for the
\textit{OLD}  than for  the \textit{YOUNG}  case, particularly  on the
negative side (see Figs \ref{dmag_g} and \ref{dmag_e}, right panels).

The  $I$  band  appears  to  be  heavily  affected  by  crowding  (see
Fig.~\ref{dmag_e}, left panel): in spite  of the low sky background in
the   optical,  the  1$\sigma$   loci  show   that  the   total  error
distributions are  very wide, and  skewed on the negative  side, again
showing the effect  of stellar blending. We notice  that the brightest
sources are affected by a  large error, which exceeds what is expected
from the shot noise. Rather than to the underlying stellar population,
we attribute this  effect to a non optimal sampling of  the PSF in the
$I$ band,  which has a  nominal FWHM of  less than 2  pixels. Although
DAOPHOT is able to reconstruct such peaked PSF, the measurement of the
magnitude  is complicated  by  uncertainties in  the  position of  the
centroid.  The   use  of  ALLFRAME,  i.e.  taking   advantage  of  the
information  on the  $J$ and  $K$  band images,  greatly improves  the
quality of the $I$  band photometry; nevertheless, an error associated
to  the under-sampling  is present  all over  the magnitude  range, and
particularly visible at the bright end.

\begin{figure}[t]
%\plotone{hb_g.eps}
\plotone{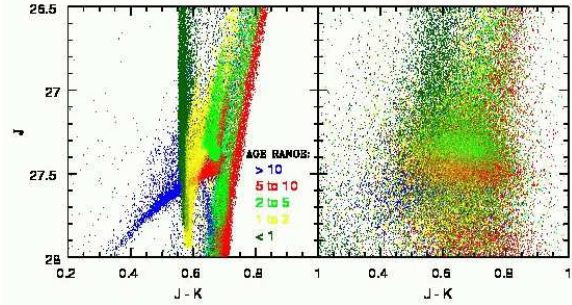}
\caption{Input (left) and output (right) CMDs in the
  region of the HB/Clump for the \textit{YOUNG}
  stellar population science case. 
  The color encodes the star's ages, as labelled.}
\label{hb_g}
\end{figure}

A science case similar to the one discussed here is presented in \cite{atul},
  albeit with some differences, including the CMD of the synthetic stellar
  population, and the exposure time, for which \cite{atul} adopt 1
  hr.  In addition, in \cite{atul}'s study, the simulated PSF has a coarser sampling
than ours, and the instrument background is neglected. In spite of these differences, for the same surface
brightness our results are broadly compatible with theirs in the $I$
and $J$ bands, with similar 1 $\sigma$ values at the bright end.

To summarize: the comparison of  the input and output catalogues shows
that, roughly for both  science cases, the output luminosity functions
become  50 $\%$  incomplete at $J,  K \sim  29$, $I  \sim 30.5$,  while a
photometric error  of $\sigma  \simeq 0.2$ mag  is reached at  $J \sim
29.5$, $K \simeq  28$ and $I \sim 30$, similar to  the S/N=5 limits in
Fig.  \ref{limits} except for  the $I$  band where  the error  is much
larger due  to crowding  \footnote{Notice that $J_{AB}  \simeq J+0.9$,
  $K_{AB} \simeq K +1.9$, $I_{AB} \simeq I + 0.1$}.  These are however
only  indicative  figures  because  the  error  distribution  is  very
asymmetrical due  to blending,  an error of  $\sigma \simeq  -0.2$ mag
being reached at  $J \sim 28.5$, $K  \simeq 27$ and $I \sim  29$. As a
result,  the output  luminosity  function is  affected  by an  overall
migration of  objects from fainter to  brighter bins, and  in order to
assess the  feasibility of a science  case it is  necessary to perform
the  whole test, proceeding,  in our  case, with  the analysis  of the
output CMDs.

\subsection{Comparison of CMDs}

\subsubsection{Disk galaxy at 4.6 Mpc} % -------------- case G ---------------------------------

Fig.~\ref{cmd_g}  shows the input  (panel (a))  and the  output (panel
(b)) CMDs for the \textit{YOUNG} stellar population science case, with
the diagnostic  boxes superimposed.  The characteristic plumes  of the
input CMD  are well recognizable in  the bright portion  of the output
diagram,   but   become    progressively   blurred   towards   fainter
magnitudes. Three  lines are  drawn on the  CMDs to  distinguish three
regimes: brighter than $J \simeq$ 26.5 the main features of the CMD in
panel (a)  are well  reproduced in panel  (b), including  their narrow
color width.   Fainter than  $K \simeq$ 28  (slanted line)  the star's
distribution on  panel (b) is completely dominated  by the photometric
errors,  and presents no  resemblance to  the corresponding  region in
panel (a).  In  the range $26.5 \leq  J \leq 28$ the CMD  in panel (b)
still retains the  main features of the corresponding  region in panel
(a), but  the color distribution of  the stars is  wider, showing that
the  photometric   errors  heavily   affect  the  appearance   of  the
CMD. Therefore, in  order to derive the SFH from  the analysis of this
region  of  the  CMD  one  needs to  quantitatively  account  for  the
photometric errors. Notice that this region includes the clump stars.

Table \ref{tab_boxes}  reports the star counts in  the four diagnostic
boxes  in the  input  and output  CMDs,  showing that  the counts  are
unaffected by  the photometric error.
Had we considered a more distant galaxy, or a stellar field
  with higher surface brightness, the diagnostic boxes would be
  closer to the confusion limit and the stellar blending would affect
  significantly the counts (starting from the RGB box,
  cf. Eq. (8) in \citealt{olsen}).  This would 
  result into a systematic error in the
  derived star formation. For the science case considered here,
  i.e. the central part of a disk galaxy at a distance of 4.6 Mpc, 
 the diagnostic boxes are bright enough that the mass transformed into stars in the
 relative age ranges can be robustly derived.
Notice that the  whole FoV of MICADO will be almost
20 times wider than simulated  here, providing many stars for a robust
statistics. 

By applying the full synthetic CMD method to the simulated
  data in Fig. \ref{cmd_g} a more detailed SFH can be
  derived, in particular disentangling the contribution of the
  various age components in the range $\sim$ 1 to 12 Gyr. To this end,
  the HB/clump region may provide an effective diagnostic.
The  left panel of Fig.  \ref{hb_g} shows the input  CMD in the
clump region with  stars color coded according to  their age bin.  The
different  colors appear  separated on  the input  CMD,  where regions
occupied by  stars in  different age bins  can be  distinguished: only
objects older than 10 Gyr are found in the blue part of the horizontal
branch;  stars with age  between 1  and 2  Gyrs populate  the faintest
envelope of  the clump,  and the typical  magnitude and color  of core
helium burning stars  is a function of the  age.  
Notice that the various
age components partially overlap, and  it is not possible to construct
clean diagnostic boxes similar to  those used in the brightest part of
the CMD.
In addition, photometric errors substantially smear out the
color distribution of the stars.
The right panel of Fig. \ref{hb_g} shows the same section
of the output  CMD with the detected objects  color coded according to
their  age. To  build this  diagram we  have assigned  to  each output
source  the age  of  its input  star  mate according  to the  position
matching criterion.   The smearing effect of the  photometric error is
readily  apparent:  the  different   age  components  of  the  stellar
population are very hard to distinguish, so that the result of
  the application of the full synthetic CMD method will be very
  sensitive to the age binning and to the photometric errors.

\subsubsection{Elliptical galaxy at Virgo }  % ------------ case E -----------------------------------

\begin{figure}
%\begin{center}%\resizebox{\hsize}{!}{\includegraphics{cmd_e_zij.eps}}
\plotone{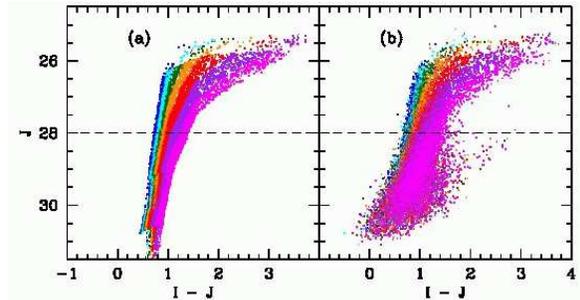}
\caption{ Input (a) and output (b) CMDs for the \textit{OLD}
  stellar population science case. The color reflects the metallicity
  bin of the object with the same encoding as on Fig. \ref{cmd_old}. The
  metallicity of the output stars is identified on the basis of the
 positional coincidence with input objects on the $J$ band image. The
 $J =28$ line is drawn to better appreciate the effect of the
 photometric errors on the width of the RGB. }
\label{cmd_e_zij}
%\end{center}
\end{figure}

Fig.  \ref{cmd_e_zij}   compares  the  input  and  output   CMDs  for  the
\textit{OLD}  stellar population  science  case. The  number of  stars
falling in the RGB box is 3538 in panel (a) to be compared to the 3616
star  counts in  the  same region  of the  CMD  in panel  (b). Due  to
blending,  some stars just  fainter than  the lower  limit of  the box
become  brighter   and  shift   inside  the  box,   contaminating  the
counts. However, the effect is quantitatively very small, and the total
mass in  stars can  be recovered using the theoretical specific  production
within few percent. Overall, the CMD of the detected sources appears very similar to
 the input CMD, especially brighter than $J \simeq 28$.
Our aim is to check the impact of the photometric errors on
a widely used method to derive the metallicity distribution of an old 
stellar population, which is based on the analysis of the CMD 
of bright RGB stars.

\begin{figure}
%\resizebox{\hsize}{!}{\includegraphics{dfeh_cmd.eps}}
\plotone{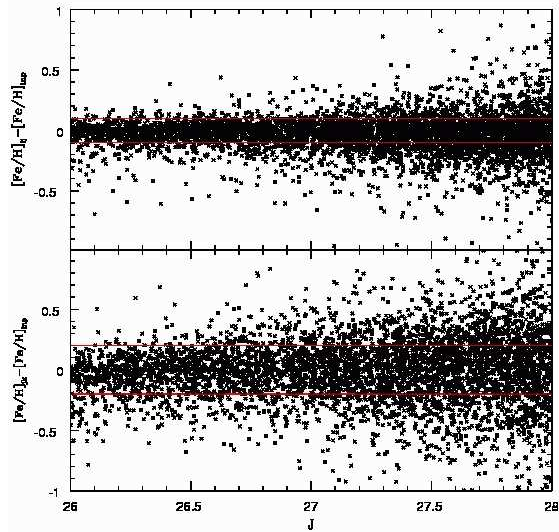}
\caption{Difference between the metallicity derived from the position
  of the detected source on the CMD and that derived from the position of the
  detected source on the frame. The upper panel shows the results from
  the analysis of the $(J,I-J)$ CMD; the lower panel from that on the $(J,J-K)$ CMD.}
\label{dfeh_cmd}
\end{figure}

\begin{figure}[t]
%\resizebox{\hsize}{!}{\includegraphics{zdist.eps}}
\plotone{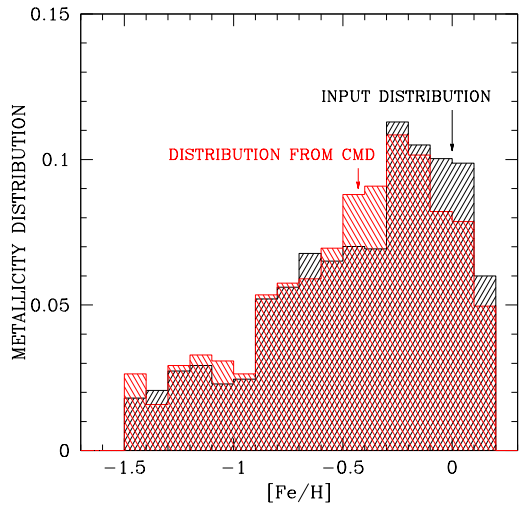}
\caption{Distribution of [Fe/H] as derived from the positional
  coincidence on the $J$ band frame compared to that derived from the
  nearest neighbor match on the $(J,I-J)$ CMD.}
\label{zdist}
\end{figure}

Fig. \ref{cmd_e_zij} shows the upper  RGB of the input (panel (a)) and
output  (panel  (b))  stellar  lists,  with  the  color  encoding  the
metallicity bin. The [Fe/H] of  each detected source is assigned equal
to  that  of  its  input  star  mate  on  the  image.   The  different
metallicity bins appear well  distinguishable on panel (b), especially
in the  upper 1.5 mag of  the RGB.  The derivation  of the metallicity
distribution from the analysis of  the CMD is subject to uncertainties
related to the age-metallicity  degeneracy and to the AGB contribution
to   the   counts   in   this    part   of   the   CMD   (see,   e.g.,
\citealt{rejkuba10}).   Besides    these   systematic   effects,   the
photometric  errors  introduce  an  additional uncertainty,  which  we
evaluate here.   We have  selected from the  output catalogue  all the
stars  brighter  than  $J =  28$,  and  assigned  to  each of  them  a
metallicity equal to that of the star in the input catalogue which has
the closest position on the CMD.  The difference between this value of
[Fe/H] and that assigned from the  source position on the image can be
viewed as  the error on the  metallicity derived from  the analysis of
the  CMD due to  mere photometric  errors.  Fig.  \ref{dfeh_cmd} shows
this difference as  a function of the output magnitude  in the J band,
the  two  panels  plotting   the  results  from  two  different  color
combinations. The  photometric errors introduce a  mild uncertainty on
the determination  of [Fe/H]  when using the  $(J,I-J)$ CMD:  for most
points the  difference between the photometrically  determined and the
true metallicity  is smaller  than 0.1 dex.  The error is  larger when
using  the  $(J,J-K)$ CMD, as anticipated in Sect. 2.2. 
Fig.  \ref{zdist} illustrates the
the impact of the errors on the photometrically derived
  metallicity distribution,
The two histograms  appear very similar: both peak at [Fe/H]
$\sim$  -0.2 and have  similar widths  and shape.  We notice  that the
photometrically derived distribution  is slightly overpopulated on the
low  metallicity   side  of  the  peak.  Overall,   however,  the  two
distributions are very close to  each other, and we conclude that this
science case  is well  feasible in inner  regions of  giant elliptical
galaxies at the distance of the Virgo Cluster.

\section{Summary and Conclusions}

In  this  paper we  have  explored  the  feasibility of  two  specific
scientific applications for  the study of the SFH  in distant galaxies
with  the expected performance  of the  42 m  E-ELT equipped  with the
MICADO  camera   working  close  to  the  diffraction   limit  of  the
telescope. We have focussed on giant galaxies located in the Centaurus
group  and in  the  Virgo Cluster.  These  are regions  of the  nearby
Universe where a variety of  galaxy types and morphology can be found,
so that  a significative sampling  of the SFH  in the Universe  can be
gathered.   In spite of  the large  collecting area  of E-ELT,  old MS
turn-offs   cannot  be  investigated   at  these   distances;  however
interesting information on the SFH can be derived from the analysis of
the luminous portion of the  CMD. The crucial advantage offered by the
E-ELT  working  close to  the  diffraction  limit  is related  to  the
exquisite  spatial  resolution  that  enables accurate  photometry  in
crowded stellar  fields. This will allow  us to study the  SFH in high
surface brightness regions of giant galaxies where most of the stellar
mass is located, and to address directly stellar population gradients,
from the outskirts down to the inner regions of galaxies.

We have investigated  how photometric errors impact on
two science cases:  A) the study of
the SFH at the center of the  disk of a spiral galaxy at a distance of
4.6 Mpc ($\mu_{\rm B} \simeq  21$); B) the recovery of the metallicity
distribution in the inner regions  (at 1/2 of the effective radius) in
an  giant elliptical  galaxy at  a distance  of 18  Mpc  ($\mu_{\rm I}
\simeq 19.6$).  To this end we  have produced synthetic  frames in the
$I, J$  and $K$ bands by distributing  theoretical stellar populations
constructed with  suitable star formation histories,  and adopting the
PSF currently  foreseen for  the MICADO camera  assisted by  the MAORY
adaptive  optics module.  The synthetic  frames were  measured  with a
standard  package  for  crowded  fields  photometry  to  yield  output
catalogues of detected stars.  The input and output stellar lists have
been  compared to determine  the photometric  quality, and  the output
CMDs have been analyzed to  assess to which extent the scientific aims
of the two cases can be realized.

For  both science  cases,  and  in  all photometric  bands,
blending   of  stellar   sources  leads   to  an   asymmetrical  error
distribution,  and  a  general  migration  of star  counts  along  the
luminosity function  towards the brighter bins. Over  a wide magnitude
range we detect an excess of stars recovered  brighter than what  they 
truly are, and the distributions of the (output--input) magnitude difference
is  more spread out  on the  negative (with  respect to  the positive)
side. This effect becomes particularly relevant as one
approaches the confusion limit, where stars are measured \textit{only} if their
luminosity is artificially boosted by blending \citep{book}.
At the same time, as the confusion limit is approached, a progressively larger
  incompleteness affects the luminosity function, partly
  balancing the excess due to blending. As a consequence, the measured luminosity
function does not appear very different from the input one (see
Figs \ref{lf_g} and \ref{lf_e}). Nevertheless the asymmetry of the error distribution persists
 and  can  lead to  misinterpretations  of the  CMD, e.g. distances
from RGB tip (and HB) stars may be underestimated; too young ages may be
inferred from Main Sequence turn off stars. Similarly, the age distribution from the core Helium
burning/clump stars could result too skewed at the young end, and
bright RGB stars overshooting the tip could be mistaken for relatively
young AGB stars. Besides the systematic shift to
  brighter magnitudes of important diagnostic features, the asymmetry of the error distribution 
affects in a systematic way the star counts in the boxes used to derive
the SFH. This effect, which is very important when trying to derive
information from stars with a magnitude close to the confusion limit,   
may also hamper the results in less severe crowding conditions.
Here we  have checked that for  our cases A)
and B),  the photometric accuracy  is sufficient for  their scientific
aims.

Specifically, our results show that in the very central parts of disks of giant
  spirals located in the Centaurus group (case A):
\begin{itemize}
\item accurate photometry ($\sigma\sim0.1$) can be obtained down to
    $J \sim 28.0$ and $K \sim 27.0$. At these
   magnitudes the luminosity functions are $\gsim$ 90 \% complete;
\item  the  characteristic plumes  of  the  input  CMD are  very  well
  reproduced in  the output CMD  down to $J  \sim 26.5$, and  the star
  counts in  bright diagnostic boxes which target  specific age ranges
  are  perfectly  reproduced. This  implies  that  the star  formation
  history in  the last $\sim$ 1  Gyr can be  accurately derived, and
  that a  robust estimate of the  mass transformed into  stars at ages
  older than $\sim 1$ Gyr can  be obtained from the star counts on the
  upper RGB;
\item the analysis  of the clump/HB stars to recover  the SFH prior to
  $\sim  1$  Gyr  ago  with  a  better  age  resolution  requires  the
  application  of the  full  synthetic CMD  method,  which includes  a
  careful modelling of the photometric errors. For the surface
    brightness considered, the smearing of the distribution of the
    clump stars is significant, hampering a robust solution.
    This case will be better performed in less crowded regions of 
    disks.
\end{itemize}

The simulations of the case B) stellar population shows that for giant
ellipticals in the Virgo cluster:

\begin{itemize}
\item accurate photometry ($\sigma\sim0.1$) can be obtained down to
  $I \sim  28.5$ and $J  \sim 27.5 $  at half of the  effective radius
  ($\mu_{\rm B}=21.6$).  At  these magnitudes the luminosity functions
  are $\gsim$ 90 \% complete;
\item the  observed CMD well  reproduces the width  of the RGB  in the
  upper 2 magnitudes,  and the color distribution of  bright RGB stars
  is only  mildly altered by  the photometric errors, 
so that its interpretation in terms of metallicity
  distribution of the stellar population is only slightly affected.
 We  estimate  that the  photometric errors  introduce an  uncertainty of
  $\simeq 0.1$ dex on [Fe/H] when using the $I-J$ color combination as
  a tracer for the effective temperature. 
\end{itemize}

Given  the good  photometric  quality  in the  bright  RGB region  for
science case B), we argue that it will be possible to study the SFH in
the  central parts  of  disks in  spiral  galaxies in  Virgo 
from star counts in
bright  boxes, as those shown  on
Fig. \ref{cmd_young}.  This  will allow us to contrast  the old (prior
to  $\sim 1$  Gyr) to  the more  recent star  formation  activity over
entire  galaxies  and  across  the  Hubble sequence  for  all  cluster
members.   The good  photometric quality  in the  upper RGB  will also
allow us to derive the  metallicity distribution almost all over giant
galaxies, thereby tracing metallicity and population gradients.

For disks  galaxies in  the Centaurus group,  the analysis of  the Red
Clump/HB feature will provide a  more detailed information on old star
formation episodes.  The central regions  of disks appear  too crowded
for this  purpose, but the  results presented here have  been obtained
with a standard package for  the photometry, which was not designed to
work with  such a structured  PSF as the  one provided by  MAORY. 
Likely, the photometric  accuracy will  improve  when using
reduction packages more suitable for complex PSF that is produced by 
adaptive optics systems.
   
Further  simulations that include a variable PSF shape over the whole
field of view and other effects (non homogeneous background, field
distortions, etc.) are required in order to fully characterize the capabilities of
ELT imaging under different conditions. At the same time, simulations
constructed with different choices for the stellar populations and
crowding conditions are necessary to adequately explore the scientific
return. These issues will be the subject of a forthcoming paper.

\appendix

\section{Procedure to generate the synthetic stars}

The  input lists  of  the  synthetic stars  were  generated by  random
extractions  of  individual  objects  which  belong to  a  library  of
theoretical simulations previously computed. For each of the two kinds
of   populations  in  Table   \ref{tab_csp},  four   simulations  were
constructed  (G.P. Bertelli,  priv. comm.),  each constrained  to have
200000   objects   brighter   than   a   given   magnitude   $M_{K,\rm
  max}=+5,+1,-2$ and  -4.  The brighter the magnitude  limit, the more
massive   the  simulated   stellar  population,   and   shorter  lived
evolutionary  phases become better  sampled. Figs  \ref{cmd_young} and
\ref{cmd_old}   show   one   of  these   \textit{mother}   simulations
respectively   for  the  \textit{YOUNG}   and  for   the  \textit{OLD}
population.  

The four  \textit{mother} simulations allow  us to construct  an input
stellar   list   for   \textit{YOUNG}  and/or   \textit{OLD}   stellar
populations of any  prescribed size, since the number  of stars in any
region of the  CMD is proportional to the total  light (or total mass)
of its parent  stellar population. In particular, the  number of stars
with magnitude in  a given range $M_{K,1} \leq  M_{K} \leq M_{K,2}$ is
proportional to the total luminosity sampled. Therefore, once the size
of the  stellar population  to be generated  is given, e.g.  its $L_B$
through  Eq. (\ref{eq_lb}),  the number  of  objects on  its CMD  with
$M_{K,1} \leq  M_{K} \leq M_{K,2}$  is determined by scaling  from the
\textit{mother}  simulation of  adequate depth,  and the  procedure is
repeated to cover  the whole magnitude range.  The  input stellar list
is thus constructed by composing partial CMDs, each containing objects
randomly extracted  from the \textit{mother}  simulations. Notice that
the  partial CMDs  are  limited in  the  $K$-band because  so are  the
original \textit{mother}  simulations. In the input  stellar list each
object  is  a  synthetic  star,  i.e.   characterized  by  mass,  age,
metallicity, and magnitude in all photometric bands.

In principle,  the input  list could contain  stars down  to $M_K=+5$,
which  is the  limit  of the  deepest  \textit{mother} simulation.  In
practice, there  is no need  to generate a  list down to  the faintest
magnitudes, which,  unless the sampled  luminosity is very  low, would
consist of an unmanageable large  number of entries. Indeed, the stars
on the synthetic  frame will be measurable only  down to some limiting
magnitude function  of the exposure time (for  the given observational
set-up).   Therefore the
stellar  population  on  the  frame  is  split  into  two  components,
i.e. individual  stars brighter than $M_{\rm K,lim}$,  plus a pedestal
of stellar light made of the stars fainter than this limit. The latter
component is evenly distributed  over the frame, with
its  Poisson noise.  

Special care has  been devoted to ensure completeness of
the list of synthetic stars  in all photometric bands, since this list
is  the  input  to  generate  frames  at  different  wavelengths.  The
extraction  procedure  described above  necessarily  generates a  list
which is complete  in the $K$ band with the  risk of under-sampling the
blue  stars.   Therefore  the  value  of $M_{\rm  K,lim}$  is  derived
imposing  completeness   in  the  bluest   band  of  the  CMD   to  be
constructed. In addition, the limiting magnitude for the input stellar
list is chosen $\sim$1.5 mag  fainter than the S/N=5 level, to account
for the effect on photometry due to stellar blending which leads to an
artificial brightening of the sources.

\acknowledgments
We are indebted  to G.P. Bertelli for computing  the synthetic stellar
populations used in our  simulations. Preliminary results of this work
are included in the Scientific Analysis Report document of the Phase A
study of the MICADO camera. We thank R. Bedin and E. Held 
for their help in an early stage of this investigation, and A. Renzini
for many useful discussions. This work was supported by funding of the
phase A study of MICADO, and by INAF TECNO funding AO@SW 1.05.01.17.06.


\begin{thebibliography}{99}
\bibitem[\protect\citeauthoryear{Aparicio et al.}{1996}]{aparicio}
  Aparicio A., Gallart C., Chiosi C., Bertelli G. 1996, ApJ, 469, L97 
\bibitem[\protect\citeauthoryear{Bertin \& Arnouts}{1996}]{Bertin96} Bertin E., Arnouts S., 1996, A\&AS, 117, 393 
\bibitem[\protect\citeauthoryear{Brown et al.}{2006}]{tom} 
Brown T.M., Smith E., Ferguson H.C., et al., 2006, ApJ, 652, 323
\bibitem[\protect\citeauthoryear{Caldwell}{2006}]{caldwell} Caldwell
  N., 2006, ApJ, 651, 822
\bibitem[\protect\citeauthoryear{Cignoni \& Tosi}{2010}]{cigno} Cignoni
  M., Tosi M., 2010, Adv. Astron., 158568
\bibitem[\protect\citeauthoryear{Cioni et al.}{2011}]{cioni} Cioni M.-R.~L., et al., 2011, A\&A, 527, A116
\bibitem[\protect\citeauthoryear{Cole et al.}{2007}]{cole} 
Cole A.A., Skillman E.D., Tolstoy E., et al., 2007, ApJ, 659, L17
\bibitem[\protect\citeauthoryear{Crnojevi\'c, Grebel \&
    Cole}{2011}]{grebel} Crnojevi\'c D., Grebel E.K., Cole A.A., 2011, A\&A, 530, A59
\bibitem[\protect\citeauthoryear {Davies \& Genzel}{2010}]{daviesmess} Davies
  R., Genzel R., 2010, The Messenger, 140, 32
\bibitem[\protect\citeauthoryear {Davies et al.}{2010}]{davies} Davies
  R. et al., 2010, SPIE, 7735, 77
\bibitem[\protect\citeauthoryear {Diolaiti et al.}{2010}]{diolaiti}
  Diolaiti E. et al., 2010, in Cl\'enet Y., Fusco T., Rousset G., eds,
  EDP Sciences, Adaptative Optics for Extremely Large Telescopes, id.02007
\bibitem[\protect\citeauthoryear {Dolphin}{2002}]{dolphin} Dolphin A. E.
 2002, MNRAS, 332, 91
\bibitem[\protect\citeauthoryear {Deep et al.}{2011}]{atul} Deep A.,
  Fiorentino G., Tolstoy E., Diolaiti E., Bellazzini M., Ciliegi P.,
  Davies R., Conan J.-M., 2011, A\&A, 531, A151
\bibitem[\protect\citeauthoryear{Falomo, Fantinel \&
    Uslenghi}{2011}]{ffu} Falomo R., Fantinel D., Uslenghi M., 2011,
  SPIE, 8135, 813523
\bibitem[\protect\citeauthoryear{Ferguson}{2007}]{annette} Ferguson
 A., 2007, in Vallenari A., Tantalo R., Portinari L., Moretti, eds, ASP Conf. Ser. Vol. 374,
 From Stars to Galaxies: Building the Pieces to Build up the Universe. Astron. Soc. Pac., San Francisco, p. 239 
%\bibitem[\protect\citeauthoryear{Ferguson et al.}{2007}]{annette}
%Ferguson A., Irwin M., Chapman S., Ibata R., Lewis G., Tanvir N., in
%de Jong R.S. ed., Dordrecht, Reidel, Island Universe - Structure and Evolution of Disk
%Galaxies, p. 239
\bibitem[\protect\citeauthoryear {Foppiani et al.}{2010}]{foppiani}
  Foppiani I. et al., 2010, in Cl'enet Y., Fusco T., Rousset G., eds,
  EDP Sciences, Adaptative Optics for Extremely Large Telescopes, id.02013
\bibitem[\protect\citeauthoryear{Gallart, Aparicio \& Vilchez}{1996}]{carma96}
  Gallart C., Aparicio A., Vilchez, J.M. 1996, AJ, 112, 1928
\bibitem[\protect\citeauthoryear{Gallart, Zoccali \& Aparicio}{2005}]{carma}
  Gallart C., Zoccali M., Aparicio A., 2005, ARA\&A, 43, 387
\bibitem[\protect\citeauthoryear{Gilmozzi \& Spyromilio}{2007}]{gilmozzi}
 Gilmozzi R., Spyromilio J., 2007, The Messenger, 127, 11
\bibitem[\protect\citeauthoryear{Girardi et al.}{2002}]{girardi}
  Girardi L., Bertelli L., Bressan A., Chiosi C., Groenewegen M.A.T.,
 Marigo P., Salasnich B., Weiss A., 2002, A\&A, 391, 195
\bibitem[\protect\citeauthoryear{Greggio et al.}{1993}]{greggio}
Greggio L., Marconi G., Tosi M., Focardi P., 1993, AJ, 105, 894 
\bibitem[\protect\citeauthoryear{Greggio}{2002}]{coimbra} Greggio
  L., 2002, in Lejune T, Fernandes J., eds, ASP Conf. Ser. Vol. 274,
  Observed HR Diagrams and Stellar Evolution. Astron. Soc. Pac., San Francisco, p. 444 
\bibitem[\protect\citeauthoryear{Greggio \& Renzini}{2011}]{book}
  Greggio L., Renzini A., 2011, Stellar Populations. A User Guide from
  Low to High Redshift., Wiley-VCH Verlag-GmbH \& Co. KGaA, Weinheim, Germany
\bibitem[\protect\citeauthoryear{Harris \& Harris}{2002}]{harris}
  Harris W.E., Harris G.L.H., 2002, AJ, 123, 3108
\bibitem[\protect\citeauthoryear{Harris \& Zaritsky}{2001}]{zari}
  Harris J., Zaritsky D., 2001, ApJS, 136, 25
\bibitem[\protect\citeauthoryear{Holtzman et al.}{1999}]{holtz}
  Holtzman J.A., Gallagher J.S.III, Cole A.A., et al., 1999, AJ, 118, 2262
\bibitem[\protect\citeauthoryear{Johns}{2008}]{johns} Johns M.,
 2008, SPIE, 6986, 3
\bibitem[\protect\citeauthoryear{Kroupa}{2001}]{kroupa}Kroupa
P., 2001, MNRAS, 322, 231
\bibitem[\protect\citeauthoryear{McQuinn et al.}{2009}]{mcq} McQuinn
  K.C.W., Skillman E.D., Cannon J.M., et al., 2009, ApJ, 695, 561
\bibitem[\protect\citeauthoryear{Olsen, Blum \& Rigaut}{2003}]{olsen} Olsen, K.A.G., Blum,
R.D., Rigaut, F., 2003, AJ, 126, 452
\bibitem[\protect\citeauthoryear{Piotto}{2009}]{piotto} Piotto 
G., 2009, in Mamajek E.E., Soderblom D.R.,  Wyse R.F.G., eds,
Proc. IAU Symp. 258, The Ages of Stars. Cambridge Univ. Press,
Cambridge, p. 233
\bibitem[\protect\citeauthoryear{Rejkuba et al.}{2005}]{rejkuba05} 
Rejkuba M., Greggio L., Harris W.E., Harris G.L.H., Peng E.W., 2005, AJ, 631, 262
\bibitem[\protect\citeauthoryear{Rejkuba et al.}{2010}]{rejkuba10} 
Rejkuba M., Harris W.E., Greggio L., Harris G.L.H., 2010, A\&A, 526, A123
\bibitem[\protect\citeauthoryear{Renzini}{1998}]{alviopix} Renzini A., 1998, AJ, 115, 2459 
\bibitem[\protect\citeauthoryear{Stetson}{1994}]{Stetson94} Stetson P.~B., 1994, PASP, 106, 250 
\bibitem[\protect\citeauthoryear{Szeto et al.}{2008}]{szeto} Szeto K.,
  et al., 2008, SPIE, 7012, 86
\bibitem[\protect\citeauthoryear{Tolstoy \& Saha}{1996}]{tolstoy} Tolstoy E.,
 Saha A., 1996, ApJ, 462, 672
\bibitem[\protect\citeauthoryear{Tosi et al.}{1991}]{tosi} Tosi M.,
 Greggio L., Marconi G., Focardi P., 1991, AJ, 102, 951
\bibitem[\protect\citeauthoryear{Weisz et al.}{2011}]{weisz} 
Weisz D.R., Dalcanton J.J., Williams B.F., et al., 2011, ApJ, 739, 5
\end{thebibliography}
\end{document}